\documentclass[aps,pre,twocolumn]{revtex4-1}

\usepackage[utf8]{inputenc}
\usepackage[T1]{fontenc}

\usepackage[normalem]{ulem}

\usepackage{amsmath}
\usepackage{amssymb}
\usepackage{amsfonts}
\usepackage{mathtools}
\usepackage{mathrsfs}
\usepackage{graphics,epsf}
\usepackage{xspace,hhline}

\usepackage{multirow}
\usepackage{calc}
\usepackage{csquotes}
\usepackage{cancel}

\usepackage[usenames,dvipsnames,svgnames,table]{xcolor}
\usepackage{graphicx}
\usepackage{filecontents}
\usepackage{lipsum}
\usepackage{tikz}
\usepackage{adjustbox}
\usepackage{braket}

\def\mean#1{\left< #1 \right>}

\newcommand*\varphib{\bar{\varphi}}
\newcommand*\phib{\bar{\phi}}
\newcommand*\psib{\bar{\psi}}
\newcommand*\Phib{\bar{\Phi}}
\newcommand*\Psib{\bar{\Psi}}
\newcommand*\etab{\bar{\eta}}

\newcommand{\la}{\langle}
\newcommand{\ra}{\rangle}

\newcommand{\p}{\partial}

\begin{document}

\title{Non-Perturbative Renormalization Group for the Diffusive Epidemic Process}

\author{Malo Tarpin$^1$, Federico Benitez$^{2,3}$, L\'eonie Canet$^1$, Nicol\'as Wschebor$^4$}
\affiliation{$^1$ LPMMC, Universit\'e Grenoble Alpes and CNRS, F-38042 Grenoble, France\\
$^2$ Physikalisches Institut, Universit\"at Bern, Sidlerstr. 5, CH-3012 Bern, Switzerland\\
$^3$ ICS, University of Zurich, Winterthurerstr. 190, CH-8057 Zurich, Switzerland\\
 $^4$ Instituto de F\'isica, Facultad de Ingenier\'ia, Universidad de la Rep\'ublica, J.H.y Reissig 565, 11000 Montevideo, Uruguay}

\begin{abstract}
We consider the Diffusive Epidemic Process (DEP), a two-species reaction-diffusion process originally proposed to model
 disease spread within a population.
This model exhibits a phase transition from an active epidemic to an absorbing state without sick individuals.
Field-theoretic analyses  suggest that this transition belongs to the universality class of
Directed Percolation with a Conserved quantity (DP-C, not to be confused with conserved-directed percolation C-DP, appearing in the study of stochastic sandpiles). However, some exact predictions derived from the symmetries of DP-C seem to be in contradiction with lattice simulations. Here we revisit the field theory of both DP-C and DEP. 
 We discuss in detail the symmetries present in the various formulations of both models. We then investigate the DP-C model using the derivative expansion of the
non-perturbative renormalization group formalism. We recover previous results for DP-C near its upper critical dimension $d_c=4$,
 but show how the corresponding fixed point seems to no longer exist below $d \lesssim 3$.
 Consequences for the DEP universality class are considered.
\end{abstract}
\maketitle

\section{Introduction}

 A large variety of systems in physics, chemistry,  biology, population genetics, or traffic flows 
 can be modeled by reaction-diffusion processes. These processes are stochastic out-of-equilibrium  models describing
 particles of one or several species which randomly diffuse on a lattice 
 and can undergo some reactions at given rates when they meet \cite{Hinrichsen00,Odor04,Marro05,Tauber05,Tauber14book}. 
 These models are of fundamental importance both from a phenomenological and from a theoretical point of view. 
 They generically feature pattern formation and non-equilibrium phase transitions, such as
  the ubiquitous transition to an absorbing state 
  which traps the system. Reaction-diffusion processes 
  hence provide simple models to study non-equilibrium scaling and phase transitions.
 Whereas the one-species processes are relatively well understood \cite{Tauber05,Tauber14book}, the multi-particle processes 
 are far less explored.

  In this work, we consider the Diffusive Epidemic Process (DEP)~\cite{Wijland98}, which is the two-species 
  ($A$ and $B$) reaction-diffusion system  with reactions:
 \begin{equation}
 A+B \xrightarrow{k} B+B, \hspace{1cm} B \xrightarrow{1/\tau} A\, . 
\label{reaction}
\end{equation}
The species $A$ and $B$ diffuse independently  with diffusion constants $D_A$ and $D_B$.
They can be interpreted as healthy and sick individuals respectively, undergoing infection on contact at rate $k$ and 
spontaneous recovery at rate $1/\tau$.
A salient feature of this process is that the total number of particles is conserved.

This model was theoretically studied using field theory and Renormalization Group (RG) in~\cite{Wijland98}.
  An action for DEP  was derived using the Doi-Peliti formalism~\cite{Doi76,Peliti85}  and  analyzed perturbatively.
For equal diffusion rates, the authors found that at a critical population density, a continuous absorbing phase 
transition occurs with upper critical dimension $d_c=4$, which does not belong to the ubiquitous Directed Percolation (DP) 
universality class
 but to a new class  (KSS) introduced in~\cite{Kree89} to model the effect of pollution on a population.
 The KSS universality class is endowed with the critical exponents $\nu=2/d$, $z=2$ and $\eta = -\epsilon/8$ 
 (that is $\beta/\nu = (d+\eta)/2 = 2-9\epsilon/16$)  to first-order in $\epsilon = 4-d$. 
For  $D_A < D_B$, the authors of~\cite{Wijland98} predicted a continuous transition of yet a new (WOH) universality 
class (distinct from DP and KSS) with exponents
 $\nu=2/d$, $z=2$ and  $\eta=0$ ($\beta/\nu=d/2$) to all orders in $\epsilon$.
For $D_A > D_B$, a fluctuation-induced first order transition was conjectured and was seemingly confirmed in~\cite{Oerding00}
 by analytical arguments
 and numerical simulations in $d=2$.
However, some of the RG predictions seem to be invalidated by further numerical simulations. The three issues concern
 i) the nature of the phase transition in the case  $D_A > D_B$,
 ii) the value of $\nu$ in the cases  $D_A < D_B$ and $D_A = D_B$,
 iii) the value of $\beta$ for $D_A \neq D_B$.

 Regarding i), all simulations performed after~\cite{Oerding00}, both in $d=1$~\cite{Fulco01,Dickman07} and $d=2$~\cite{Dickman08},
 strongly indicate that the phase transition is continuous also in the case $D_A > D_B$, checking in particular the absence of 
 hysteresis~\cite{Dickman08}.
 Regarding ii),  early simulations in $d=1$ for $D_A=D_B$~\cite{Freitas00} found $\nu=2.21(5)$ in disagreement with the RG prediction 
 $\nu=2/d=2$~\cite{Kree89,Wijland98}. Subsequent simulations reported in~\cite{Fulco01} partially reconcile both results suggesting that
  the discrepancy
 could be imputed to corrections to scaling.  However, the debate exposed in~\cite{Janssen01,Freitas01} is still unresolved. 
The initial result $\nu=2.21(5)$ of~\cite{Freitas00} was criticized by Janssen~\cite{Janssen01} using the following line of argument. By truncating the action
deduced in~\cite{Wijland98} to the terms relevant in a perturbative analysis around the upper critical dimension, he obtained an action
which was given the name of DP-C~\footnote{The sandpile community use the name C-DP (for 
 \textit{conserved-directed percolation}) to refer to the specific case $D_A=0$, see~\cite{Janssen16}}.
The DP-C action exhibits particular 
  symmetries which enforce
 the exact value $\nu=2/d$ at the corresponding fixed point. On the other hand, the authors of~\cite{Freitas01} replied by observing that
 the full DEP action  includes (irrelevant) terms that violate
 these symmetries. Although these terms are naively irrelevant near the upper critical dimension $d_c=4$, 
 they could become relevant away from it and in particular in $d=1$.
 If the transition is not driven by the DP-C fixed point, but instead by another one  having less symmetries, the argument of~\cite{Janssen01} does not hold and the value of $\nu$ is not fixed. It could depart from $2/d$ and possibly be compatible with values
 from simulations. Numerical simulations in $d=2$ for equal diffusion constants $D_A=D_B$ convincingly ruled out the DP exponents,
  but could not settle on
 whether $\nu=1$ (with possible logarithmic correction) in accordance with the field-theoretic result, or $\nu<1$~\cite{Bertrand07}.
 Finally, regarding iii), Table~\ref{tab1} shows that for $D_A<D_B$ in $d=1$, either $\nu \neq 2$ or $\beta \neq 1$, and likewise 
 for $D_A>D_B$. Yet,  if the transition is controlled by the DP-C fixed point, the symmetries constraints
 imply $\nu = 2/d=2$  and $\beta = 1$.

In this work we present our take on these issues  along two complementary lines.
First, we review  the different actions proposed both for DEP and DP-C, and perform 
 a detailed analysis of the symmetries of these actions. We show that the action of DP-C 
  has more symmetries than that of DEP. These additional symmetries are likely to be realized in 
   the neighbourhood of the upper critical dimension $d_c$.
   The question, already raised in \cite{Janssen01,Wijland98}, is whether 
  they can effectively emerge in the infrared near the critical regime far from $d_c$.
  Accordingly, in a second step, we study the DP-C action to search
  for an infrared fixed point in low dimension, using the non-perturbative renormalization group (NPRG) formalism~\cite{Wetterich93,Tetradis94} at leading order
in the Derivative Expansion approximation scheme. This approximation scheme  has proven
  to be generically very effective to tackle
difficult problems  beyond perturbation theory (for  general reviews, see~\cite{Bagnuls01,Berges02,Kopietz10}; for a pedagogical introduction, see~\cite{Delamotte12}).
The application of the NPRG method to reaction-diffusion processes was initiated in~\cite{Canet04} and has a
fruitful history for one-species processes (see, for example, \cite{Canet04-2,Canet05,Gredat12,Benitez12,Bartels15}; for a recent technical improvement, see \cite{Duclut16}). 
It is here implemented for the first time to study a multi-species reaction-diffusion system.
 Using this method, we observe that the fixed point found in perturbation theory acquires an additional  
 relevant direction at a dimension
around $3$. This suggests that the DP-C fixed point does not control the critical physics of the DEP universality class in dimensions
 $d=1$ and $d=2$. We leave for future work the analysis of a more general approximation scheme accounting for the terms
  which are allowed by the symmetries of the DEP action but forbidden by the more restrictive DP-C one. 
 We finally present our conclusions and some perspectives.
 
\begin{table}
\begin{tabular}{||c|c|c|c|c|c|l||}
\hline\hline
\;\;\; $d$\;\;\; &\;\;\;$\mu$\;\;\;&  $\beta/\nu$ & $\beta$ & $\nu$ & $z$ & Ref.\\
\hline
\hline
     &       &  0.197(2) & 0.435(14) & 2.21(5) &  --     &~\cite{Freitas01}\\ 
     &   0   &  0.226(20)& {\color{red} 0.452(40)} & --  &  -- &~\cite{Fulco01}\\ 
     &       &  0.192(4) & 0.384(46) & 2.0(2) & 2.02(4) &~\cite{Dickman07}\\
     &       &   --   & -- & 2.037  &  1.980  &~\cite{Filho10} \\
     &       &  \textit{0.3125} &  & \textit{2} & \textit{2} &~\cite{Wijland98,Janssen01}\\
\hhline{||~|-|-|-|-|-|}
     &       &  0.113(8) & 0.20(2) & 1.77(3) & 1.6(2)  &~\cite{Dickman07}\\ 
1    & $>$0  &  0.165(22)& {\color{red} 0.330(44)} & -- & -- &~\cite{Fulco01}\\ 
     &       &   --      & -- & 2.0  &  1.992  &~\cite{Filho10} \\
\hhline{||~|~|-|-|-|-|~|}
     &       &  \multicolumn{4}{c|}{\textit{first order}}    &~\cite{Wijland98}\\
\hhline{||~|-|-|-|-|-|}
     &       & 0.404(10) & 0.929(144) & 2.3(3) & 2.01(4) &~\cite{Dickman07}\\
     &  $<$0 &  0.336(15)& {\color{red} 0.672(30)} & -- & --      &~\cite{Fulco01}\\ 
     &      &   --      & -- & 2.0 & 1.992  &~\cite{Filho10} \\
     &       & $\mathit{1/2}$ & \textit{1} & \textit{2} & \textit{2}       &~\cite{Wijland98,Janssen01}\\  
\hline
\hline
     & 0 & 0.856(4) & 0.797(8) & 0.932(5)   & -- &~\cite{Bertrand07}\\
2    &       &  \textit{0.875} &  & \textit{1} & \textit{2}       &~\cite{Wijland98,Janssen01}\\
\hhline{||~|-|-|-|-|-|}
    & $>$0 & 0.88(5) & 0.93(9) & 1.06(4) & 1.89(8) &~\cite{Dickman08} \\
    \hhline{||~|~|-|-|-|-|~|}
     &  &  \multicolumn{4}{c|}{\textit{first order}}    &~\cite{Wijland98}\\

\hline\hline
\end{tabular}
\caption{Critical exponents of DEP from Monte Carlo simulations and field theoretical analyses. The values in gray 
(red online) are  deduced assuming $\nu = 2/d$ and the values in italic are theoretical predictions. The parameter $\mu$ is  defined as $\mu=(D_A-D_B)/D_A$. }                                                                             
\label{tab1}
\end{table}

\section{Field theory and Doi-Peliti procedure}

A process such as (\ref{reaction}) defines a master equation. Following the Doi-Peliti procedure~\cite{Doi76,Peliti85}, the latter can be cast into a field theory
 upon introducing creation and annihilation operators and formulating a coherent-state path integral representation for the mean value of
 any function $O$ of the density of the different species in the model. For example, for processes with two species $A$ and $B$, $O$ is a function
 of the densities of $A$ and $B$ (respectively $n_A$ and $n_B$):
  \begin{equation}
  \mean{O(n_A,n_B)} = \mathcal{N}^{-1} \int \mathcal{D}[a,a^*\negmedspace,b,b^*] O(a,b)e^{-S [a,a^*\negmedspace,b,b^*]}\,.
 \end{equation}
In particular, for the DEP process (\ref{reaction}) one obtains
 \begin{align}
 S_{\rm DEP} [a,a^*\negmedspace,b,b^*] = \int_{\textbf{x},t} \Big\{ a^*(\partial_t-D_A\Delta)a + b^*(\partial_t-D_B\Delta)b \nonumber\\
  + k a b (a^*-b^*)b^* - \frac{1}{\tau} b (a^*-b^*)\Big\}\, ,
  \label{DEP}
 \end{align}
 with  $a^* \text{ (resp. }b^*)$ complex-conjugate of $a\text{ (resp. }b)$, and $n_A=a^* a$, $n_B=b^*b$.
 Since we are  interested only in stationary-state expectation values, the dependence with respect to initial conditions 
 is dropped  in the previous expressions, and in the following. 
 
One can define, as usual, the generating functional for correlation functions:
\begin{equation}
\label{partfunc}
 Z[J] \equiv \int \mathcal{D}[\phi] \, e^{-S[\phi] + J^{\alpha}\phi_{\alpha}}\, ,
\end{equation}
where greek indices stand for coordinate and internal indices. Here, $\phi$ denote the various fields present in the model and $J$ the associated sources.
 The cumulant generating functional and its Legendre transform (the generating functional of one-particle-irreducible  vertex functions, or
  effective action $\Gamma[\Phi]$) are defined as:
\begin{align}
 W[J] &\equiv \log \,Z[J]\\
 W[J] + \Gamma[\Phi] &\equiv J^{\alpha}\Phi_{\alpha}\, ,
\end{align}
where $\Phi_{\alpha} \equiv \mean{\phi_{\alpha}} = {\delta W}/{\delta J_{\alpha} }$. 

Working with the effective action has the advantage  that the symmetries of the original action are expressed
   as simple Ward identities for the effective action.
More precisely, let us consider infinitesimal symmetries that are affine in the fields:
\begin{equation}
\label{Svar}
 \phi_{\alpha} \to \phi_{\alpha} + \delta_{\epsilon}\phi_{\alpha},\quad \delta_{\epsilon}\phi_{\alpha} = \epsilon(A^{\beta}_{\alpha}\phi_{\beta}+B_{\alpha}) \, .
\end{equation}
The related Ward identity imposes that $\Gamma[\Phi]$ possesses the same symmetry, that is, $\Gamma[\Phi]$ verifies
\begin{equation}
 \delta_{\epsilon}\Gamma[\Phi]  = \frac{\delta \Gamma}{\delta \Phi_{\alpha}}\delta_{\epsilon}\Phi_{\alpha} = 0 \, .
\end{equation}
This result generalizes to
 the case where the action $S[\phi]$ is not strictly invariant under the transformations (\ref{Svar}), but has a variation 
  {\it linear in the fields}. In this case, the Ward identity simply becomes 
\begin{equation}
 \delta_{\epsilon}\Gamma[\Phi]  = \frac{\delta \Gamma}{\delta \Phi_{\alpha}} \delta_{\epsilon}\Phi_{\alpha} = \langle\delta_{\epsilon}S\rangle =\delta_{\epsilon}S[\Phi]\, ,
\end{equation}
that is, the variation of the effective action has the same form as the variation of the bare action.
We now analyze the various actions proposed for the study of the transition in the DEP model and the associated symmetries.

\section{DEP field theory}

In order to study the action (\ref{DEP}) and following~\cite{Wijland98}, we apply the Doi-shift and use the change of variables
\begin{gather}
 \varphi = \rho^{-\frac{1}{2}}(a + b - \rho),\quad \psi = \rho^{-\frac{1}{2}} b,\nonumber\\
 \varphib = \rho^{\frac{1}{2}}\tilde{a},\quad \psib = \rho^{\frac{1}{2}}\,(\tilde{b}-\tilde{a})\, ,
\end{gather}
where $\rho$ is the initial total density, to obtain the action of DEP~\cite{Wijland98}
\begin{align}
 S_{\text{DEP}}^{\rm WOH} = \int_{\textbf{x},t}\Big\{ \varphib(\partial_t-\Delta)\varphi+\psib(\partial_t -\lambda\,\Delta - \sigma)\psi\nonumber\\
  +\mu\,\varphib\Delta\psi+\psi\psib \Big[g(\psi-\psib - \varphi -\varphib)\nonumber\\
  +v(\psi\psib + \psi\varphib-\psib\varphi-\varphi\varphib)\Big]\Big\}\, ,
  \label{CSPI}
\end{align}
with parameters
\begin{gather}
  g = \frac{k\sqrt{\rho}}{D_A},\quad v = \frac{k}{D_A},\quad \sigma = k(\rho-\rho_c)/D_A,\quad\rho_c = (k\tau)^{-1}\nonumber\\
  \lambda = D_B/D_A,\quad \mu = (D_A-D_B)/D_A \, .
\end{gather}
In the  expression (\ref{CSPI}), the time has been rescaled by $D_A$. As a consequence, we cannot study directly the case $D_A=0$,   known as C-DP in the literature,
which is a whole field in itself~\cite{Manna91, PastorSartorras00, Dickman01,Hoai10, Wiese16}. Notably, a mapping has been established between C-DP and the quenched
Edward-Wilkinson model in~\cite{LeDoussal15, Janssen16}. It would be interesting to investigate whether the limit $D_A \to 0$ can be  recovered from the DEP action as a short time transitional regime, since this limit corresponds to $t\to 0$ in the rescaled action.
 This would imply to consider breaking of time-translation invariance, which is beyond the scope of the present work.

Further approximations of the DEP action are delicate because the fields in (\ref{CSPI}) do not represent the physical densities. 
In particular, we show below that the truncation
 of $S_{\text{DEP}}^{\rm  WOH}$ performed in~\cite{Wijland98} leads to an action which does not conserve the total number 
 of particles if interpreted as a coherent-state path integral.
 To clarify this point, let us  expound an alternative way to obtain a coarse-grained action~\cite{Janssen16}  which relies on the
Grassberger transformation~\cite{Grassberger82}
 \begin{equation}
  a^* = e^{\tilde{n}_A},\quad b^* = e^{\tilde{n}_b},\quad a = n_A e^{-\tilde{n}_A},\quad b = n_B e^{-\tilde{n}_B}\, .
 \end{equation}
It was shown in~\cite{Andreanov06} that this transformation results in a path integral for the physical density fields of particles. 
 Applied to  the original action (\ref{DEP}), it yields
 \begin{align}
  S_{\text{DEP}}^{\rm G} = \int_{\textbf{x},t}&\Big\{\tilde{n}_A(\partial_t-D_A\Delta)n_A - D_A n_A \big( \nabla \tilde{n}_A \big)^2\nonumber\\
  &+ \tilde{n}_B(\partial_t-D_B\Delta)n_B - D_B n_B \big( \nabla \tilde{n}_B \big)^2\nonumber\\
  &+ k n_A n_B \big(1 - \exp(\tilde{n}_B - \tilde{n}_A)\big) \nonumber\\
  &- \frac{1}{\tau} n_B \big(\exp(\tilde{n}_A - \tilde{n}_B) - 1\big)\Big\}\, .
  \label{Grassberger}
 \end{align}
In this formulation,
the conservation of the total number of particles is conveniently expressed  through the following
 field transformation~\cite{Janssen16}:
\begin{equation}
\tilde{n}_A \to \tilde{n}_A + \Lambda(t),\quad\tilde{n}_B \to \tilde{n}_B +\Lambda(t)\, ,
 \label{symI}
\end{equation}
where $\Lambda(t)$ is  a function of time with suitable boundary conditions.
 Even if this transformation is not a symmetry of the action 
$S_{\text{DEP}}^{\rm G}$ given by (\ref{Grassberger}), the variation of  the latter
is linear in the fields. As explained in the previous section, one can then deduce simple Ward identities. 

Let us come back to the coherent-state action of DEP, Eq. (\ref{CSPI}). Although the conservation of particles is not as simple
 as in the Grassberger formulation (transformation (\ref{symI})), the corresponding symmetry exists for the action (\ref{CSPI}) and reads
\begin{gather}
      \quad \varphib \to \Lambda \varphib + (\Lambda-1)\frac{g}{v}\nonumber\\
			\quad \varphi \to \Lambda^{-1} \varphi + (\Lambda^{-1}-1)\frac{g}{v}\nonumber\\
                    \quad \psib \to \Lambda \psib, \quad \psi \to \Lambda^{-1} \psi \, ,
    \label{symIII}
\end{gather}
where $\Lambda$ is an arbitrary constant.
Let us remark that the field transformation (\ref{symIII}) is a true symmetry of the action (\ref{CSPI}),  contrarily to its Grassberger version, 
 where the transformation (\ref{symI}) with a constant $\Lambda$ is not a symmetry of $S_{\text{DEP}}^{\rm G}$. 
   On the contrary, the Ward identity associated with  the time-gauged transformation (\ref{symIII}) turns out to be difficult to control 
 in the WOH formulation (because in this case the variation of the action is not linear in the fields).
 On the other hand, the action $S_{\text{DEP}}^{\rm WOH}$ is a polynomial in the fields
 which significantly simplifies most calculations. Let us emphasize that, in both cases,  truncations of the corresponding actions
  -- which are necessary in a perturbative approach -- must be done carefully, in order not to break the symmetries (\ref{symI}) or (\ref{symIII}),
 associated with the conservation of the total number of particles.
In the next section, we analyze the action for the DP-C model and discuss its relations and differences with the DEP action.

\section{DP-C field theory}

Despite of the transparent symmetry (\ref{symI}), the Grassberger action (\ref{Grassberger}) has the drawback of being
 no longer a polynomial in the fields. Nonetheless, by further heuristic truncations, one can propose the action of the \textit{Directed Percolation coupled to a Conserved field} or DP-C. 
 From (\ref{Grassberger}), changing to variables 
 $ n_A \to \rho + c - n, n_B \to n, \tilde{n}_A \to \tilde{c}, \tilde{n}_B \to \tilde{n} + \tilde{c}$,
  keeping only relevant terms around
 the upper critical dimension $d_c$ and rescaling, one gets
 \begin{align}
  S_{\text{DP-C}}^{\rm G} = \int_{\textbf{x},t}\Big\{ \tilde{c}(\partial_t-\Delta)c +\tilde{n}(\partial_t-\lambda\,\Delta - \sigma)n \nonumber\\
   + r\, \tilde{c}\Delta\tilde{c} + \mu\, \tilde{c}\Delta n
  + \gamma\, n \tilde{n}(n-\tilde{n}-c)\Big\}\, ,
  \label{DP-C}
 \end{align}
using the same notations as before and with
\begin{equation}
  \gamma = \frac{k}{D_A}\sqrt{\frac{\rho+\rho_c}{2}},\, r = \frac{2\rho}{\rho + \rho_c}\, .
\end{equation}
In this action, $n$ represents the density of B, equivalent to $\psi$, and $c$ the fluctuation of the total density, equivalent to $\phi$.
It is an extension for arbitrary $\mu$ of the KSS action obtained in~\cite{Kree89} from a coarse-grained Langevin equation.

The DP-C action can also be obtained from the coherent-state action (\ref{CSPI}), upon dropping the quartic terms which
 are perturbatively irrelevant in $d=4-\epsilon$. This was done 
in~\cite{Wijland98} and led for $\mu <0$ to the WOH universality class. However, the resulting action interpreted as a coherent-state path integral does not correspond to a process where the total number of particles is conserved. This can be checked by inverting the changes of variables to re-express
  the action in terms of  $a,a^*,b,b^*$, and then applying the Grassberger transformation.
 This issue notwithstanding, the truncated action is formally equivalent to DP-C. Indeed using the change of variable 
 $c = \varphi+\varphib,\,\tilde{c}=\varphib$, one recovers (\ref{DP-C}) with $\gamma=g$ and $r=1$. (Note that in (\ref{DP-C}), $r$ can be absorbed by a rescaling of the  fields $c$ and $\tilde c$).
Thus, at the formal level of the field theory, forgetting about the nature of the fields at stake, the density field action and the
coherent-state one are equivalent when they are  truncated to the terms  relevant around the upper critical dimension.
In the remainder of the article, we choose to work with the latter, which reads:
\begin{align}
  S_{\text{DP-C}}^{\rm WOH} = \int_{\textbf{x},t}\Big\{ \varphib(\partial_t-\Delta)\varphi+\psib(\partial_t -\lambda\,\Delta - \sigma)\psi\nonumber\\
  +\mu\,\varphib\Delta\psi+g\,\psi\psib (\psi-\psib - \varphi -\varphib)\Big\}\, .
  \label{DP-C-Doi}
\end{align}
The shift transformation (\ref{symI})  in terms of these fields reads 
\begin{equation}
 \varphi \to \varphi + \Lambda(t),\quad \varphib \to \varphib - \Lambda(t) \,,
 \label{symI'}
\end{equation}
with $\Lambda(t)$ a function of $t$ with appropriate boundary conditions as before.
  As for the Grassberger action (\ref{Grassberger}), the variation of the DP-C action
(\ref{DP-C-Doi}) under the transformation (\ref{symI'}) is linear in the fields, generating simple Ward identities. Even more,
 for the truncated action (\ref{DP-C-Doi}), one can consider the generalization of (\ref{symI'}) gauged not only in time but also in space:
\begin{equation}
 \varphi  \to \varphi + \Lambda({\bf x},t),\quad \varphib \to \varphib - \Lambda({\bf x},t) \, .
 \label{spacetimegauged}
\end{equation}
The variation of the DP-C action under this enhanced transformation is linear in the fields, which implies the Ward identity:
\begin{equation}
\label{wardidsansreg}
 \delta_{\epsilon}\Gamma = \delta_{\epsilon}\Big\{\Phib(\partial_t-\Delta)\Phi + \mu\,\Phib\Delta\Psi\Big\}\, ,
\end{equation}
where $\delta_{\epsilon}X$ denotes the variation of $X$ under the infinitesimal transformation (\ref{spacetimegauged}). This prompts to write
\begin{equation}
 \Gamma = \int_{\textbf{x},t} \Phib(\partial_t-\Delta)\Phi + \mu\,\Phib\Delta\Psi + \tilde{\Gamma}[\Phi+\Phib,\Psi,\Psib]\, ,
\label{gammagen}
\end{equation}
{\it i.e.} apart from the first two terms which are not renormalized, $\Gamma$ is invariant under (\ref{spacetimegauged}).
This result implies that $d_{\Phi} = d_{\Phib} = d/2$ (which  prompts to define $\eta$ and $\etab$ as the anomalous dimension of $\Psi$
 and $\Psib$ respectively) and $z = 2$. Moreover if $\mu \neq 0$, then $\Psi$ and $\Phi$ must have the same dimension, 
  which leads to $d_{\Psi}=d/2$, {\it i.e.} $\eta = 0$.
 
Besides this gauged shift transformation, the DP-C action has the following duality property \cite{Janssen01},
\begin{equation}
\varphi \to \varphi + \epsilon,\quad \sigma \to \sigma - g \epsilon\, ,
 \label{symII}
\end{equation}
where $\epsilon$ is a constant real parameter. It was shown in \cite{Kree89,Janssen01} that this duality implies $\nu = 2/d$ exactly.
  This result, in conjunction with $\eta=0$ (which holds for $\mu\neq 0$), then yields $\beta = 1$ exactly  through the
 hyperscaling relation $\beta/\nu = (d+\eta)/2$. 
In the particular case where $\mu=0$, the action (\ref{DP-C-Doi}) possesses the extra generalized  ``rapidity'' symmetry \cite{Janssen01}:
\begin{align}
\psi({\bf x},t)& \to -\psib({\bf x},-t),\quad \psib({\bf x},t) \to -\psi({\bf x},-t), \nonumber\\
\varphi({\bf x},t)& \to \varphib({\bf x},-t), \quad \varphib({\bf x},t) \to \varphi({\bf x},-t)\, .
\end{align}
As a consequence, the effective action $\Gamma$ inherits this symmetry and, moreover, $\eta=\etab$.
All the exact results for the exponents of the DP-C universality class (equivalent to the KSS class for $\mu=0$ and 
 to the WOH class for $\mu < 0$)  are summarized in Table~\ref{exact}.
\begin{table}
\begin{tabular}{| c | c | c |}
\hline
\hline
 \quad\quad\quad \quad\quad    &\quad\quad  $\mu = 0$ \quad\quad   & \quad\quad $\mu \neq 0$ \quad\quad 	 \\
\hline
$\nu$  & $2/d$ & $2/d$ \\
$z$ & 2 & 2 \\
$\eta$ & $\etab$  & 0 \\
$\beta$  &       & 1\\
\hline
\hline
\end{tabular}
\caption{Exact critical exponents of the DP-C universality class.}
\label{exact}
\end{table}

Let us make a point clear about issue iii) in the introduction, regarding the value of the exponent $\beta$.
  One of the outcomes of the previous analysis is that for DP-C, the identities $\nu = 2/d$ and $\eta = 0$
   imply $\beta = 1$ exactly for $\mu \neq 0$. 
  Hence the DP-C action cannot account for a non-trivial value of $\beta$, as seems to be  observed in this case
   in lattice simulations~\cite{Fulco01, Dickman07}.

\section{Discussion of the relation between DEP and DP-C}
\label{scenario}

The symmetry (\ref{symIII}) provides a possible scenario for the transition from DEP to DP-C. 
Indeed, the corresponding Ward identity, derived in (\ref{ward-DEP}), reads for arbitrary vertex functions:
\begin{align}
\Big\{\sqrt{\rho}\Big(\frac{\partial}{\partial \phi_0} - \frac{\partial}{\partial \phib_0}\Big) 
+ \phi_0 \frac{\partial}{\partial \phi_0} - \phib_0\frac{\partial}{\partial \phib_0} + \psi_0 \frac{\partial}{\partial \psi_0} - \psib_0 \frac{\partial}{\partial \psib_0} \nonumber\\
+ (a-b+c-d)\Big\}\Gamma^{(a,b,c,d)}\big(\{\textbf{q}_i,\omega_i\};\phi_0,\phib_0,\psi_0,\psib_0\big)= 0 \, ,
\end{align}
where the subscript $0$ denotes constant and uniform background fields.
 The functional $\Gamma$ is defined at a renormalization scale $k$ which goes to zero in the thermodynamic limit (see next section).
 Let us introduce the renormalised adimensioned fields $\hat \phi = k^{d_\phi} \phi$, $\hat {\phib} = k^{d_{\phib}} \phib$,  $\hat \psi = k^{d_\psi} \psi$, $\hat {\psib} = k^{d_{\psib}} \psib$. 
In the corresponding adimensioned identity, the first two terms are enhanced with an explicit $k^{-d_\phi}$ and  $k^{-d_{\phib}}$ factor. Hence,  they dominate in the limit $k \to 0$, as long as the scaling dimensions of the fields $\phi$ and $\phib$ are positive. One is then left in the infrared with the following Ward identity
\begin{equation}
 \Big(\frac{\partial}{\partial \phi_0} - 
 \frac{\partial}{\partial \phib_0}\Big)\Gamma^{(a,b,c,d)}\big(\{\textbf{q}_i,\omega_i\};\phi_0,\phib_0,\psi_0,\psib_0\big)= 0 \, ,
\label{wardshiftGamma}
\end{equation}
 which is identical to the Ward identity for the pure shift (\ref{symI'}).
In fact, we present below results from the
integration of the NPRG flow which confirm to some extent this scenario. 

However, let us note that this property is not sufficient to ensure that the DEP transition belongs to the universality class of DP-C.
 Indeed, the exponents of DP-C follow from the conjunction of both the space-time gauged shift (\ref{symI'}) and the duality relation 
(\ref{symII}). However, the first of these symmetries is not valid for DEP, since a renormalization of the $\phi\phib$ propagator 
 is allowed. Consequently $\nu$ can still depart from $2/d$ even though the shift symmetry is restored in  the infrared at the DEP fixed point.

 Of course, these symmetries can always appear as accidental symmetries of the fixed point.
  Indeed, the field theoretical results \cite{Wijland98,Janssen01} suggest that this is the case for spacial
 dimensions close to the upper critical dimension, where the only relevant terms are the cubic ones,
  and the cubic action satisfies these gauged symmetries. 
 However, it is unclear whether the gauged identities remain attractive at the relevant infrared fixed point in arbitrary dimensions.
 This connects with the original question raised in \cite{Wijland98,Janssen01}: The DP-C fixed point has {\it a priori} more symmetries 
 than the DEP one. Whether the DP-C fixed point continues to control the DEP transition in low
 dimensions (as it does near the upper critical dimension), is a dynamical question to be investigated.
  Our task in the remainder of the article is  to study both DP-C and DEP through their respective symmetries with the help of NPRG techniques.

\section{Non-perturbative renormalization group}

 The NPRG procedure for equilibrium systems \cite{Wetterich93,Tetradis94} 
 can be straightforwardly generalized to out-of-equilibrium
phenomena~\cite{Canet04, Canet11}. 
The starting point  is to modify the generating functional (\ref{partfunc}) by adding a scale dependent
quadratic regulator. Generically,
\begin{equation}
 Z_k[J] \equiv \int \mathcal{D}[\phi] \, e^{-S -\frac{1}{2}R_k^{\alpha\beta}\phi_{\alpha}\phi_{\beta} + J^{\alpha}\phi_{\alpha}}\, ,
\end{equation}
where, as before, greek indices stand for coordinate and internal indices. 
The role of the regulator $R_k$ is to  smoothly replace momenta $q < k$ in the propagator with $k$ (denoting $q=|\textbf{q}|$). Here, both the 
$\varphib\varphi$ and  $\psib\psi$ propagators are regulated; more precisely
\begin{align}
\label{reg}
 \frac{1}{2}R_k^{\alpha\beta}\phi_{\alpha}\phi_{\beta} = \int_{\textbf{q},\omega} R_k^{\varphi}(q) \varphib(-\textbf{q},\,\omega)\varphi(\textbf{q},\,\omega) \nonumber\\
 + R_k^{\psi}(q) \psib(-\textbf{q},\,\omega)\psi(\textbf{q},\,\omega)\, ,
\end{align}
with $R_k^{\varphi}(q) = q^2 r(\frac{q^2}{k^2})$ and $R_k^{\psi}(q) =\lambda_k Z_k q^2 r(\frac{q^2}{k^2})$,
where $\lambda_k$ and $Z_k$ are running coefficients defined in Eq. (\ref{ansatzDPC}).
In this work, we choose to use the   cutoff function $r$ proposed by Litim~\cite{Litim01}
\begin{equation}
\label{Theta}
r(y) = \big(\frac{1}{y}-1\big)\Theta(1-y)\, ,
\end{equation}
 which allows for analytical integration on momenta. The regulated cumulant generating functional and its (modified) Legendre transform
 are then defined as
\begin{align}
 W_k &\equiv \log \,Z_k\\
 W_k + \Gamma_k &\equiv J^{\alpha}\Phi_{\alpha} - \frac{1}{2}R_k^{\alpha\beta}\Phi_{\alpha}\Phi_{\beta}\, ,
\end{align}
where $\Phi_{\alpha} = \mean{\phi_{\alpha}} =  \delta W_k/\delta J_{\alpha}$,
 generalizing previous definitions in presence of the infrared regulator.
The evolution of $\Gamma_k$ with the RG scale $k$
 is given by the exact flow equation ~\cite{Wetterich93}:
\begin{equation}
  \partial_k \Gamma_k = \frac{1}{2} \partial_k R_k^{\alpha\beta}(\Gamma_k^{(2)}+R_k)^{-1}_{\beta\alpha}\, ,
  \label{Wetterich}
\end{equation}
which smoothly interpolates between the mean-field effective action $\Gamma_\Lambda\sim S$ and the exact effective action
 $\Gamma_0 = \Gamma$, when $k$ is
lowered from a microscopic UV cutoff $\Lambda$ to $0$. At this point, in order to integrate (\ref{Wetterich}), 
some approximations have to be made based on the properties of $R_k$ and on identities coming from the symmetries of $\Gamma_k$.

\subsection{Exact exponents from symmetries of DP-C}

In order to propose appropriate ans\"atze for $\Gamma_k$, its general form must be constrained by the symmetries of the action.
This is achieved as previously through the use of Ward identities, generalized to account for the presence of the infrared regulator $R_k$
 (see Appendix~\ref{ward}). 
  In particular, the Ward identity (\ref{wardidsansreg}) remains true for $\Gamma_k$, 
 and yields that  $\Gamma_k$ endows the same general form as (\ref{gammagen})
\begin{equation}
 \Gamma_k = \int_{\textbf{x},t} \Phib(\partial_t-\Delta)\Phi + \mu\,\Phib\Delta\Psi + \tilde{\Gamma}_k[\Phi+\Phib,\Psi,\Psib]\, ,
\end{equation}
with the explicit kinetic terms unrenormalized and $\tilde\Gamma_k$  invariant under (\ref{symI'}).
This implies that the NPRG flow maintains the property $d_{\Phi} = d_{\Phib} = d/2$ and $z = 2$.
Moreover, if $\mu \neq 0$, then the flow also preserves $d_{\Psi}=d/2$,  {\it i.e.} $\eta = 0$.
 
Furthermore, using the duality transformation (\ref{symII}), one obtains the constraint (see Appendix \ref{ward-dpc})
\begin{equation}
 g\,\p_{\sigma} U_k = \p_{\Phi} U_k\, ,
 \label{ward0}
\end{equation}
where $U_k$ is the effective potential, which  corresponds to $\Gamma_k$ evaluated for homogeneous static fields 
(divided by the space-time volume). Within an adequate parametrization of $U_k$, one can deduce from (\ref{ward0})
 that $\nu = 2/d$ exactly, and at the same time identify the relevant direction of the flow, see Appendix~\ref{ward}.
  Hence, within the NPRG framework, the modified Ward identities   
 also fix the exact values for the critical exponents of DP-C, gathered in Table~\ref{exact}. Let us notice that for DEP, a relation similar to (\ref{ward0}) also exists (see Appendix \ref{ward-DEP}), 
  fixing $\nu = 1/d_{\phi}$ exactly. However, the field $\varphi$ may
 acquire in this case a non-trivial dimension so that $\nu$ can be different from $2/d$.
 
\subsection{$\Gamma_k$ ansatz for DP-C and results from NPRG}

Using the above contraints, we propose an ansatz for $\Gamma_k$. In this work, we restrict to the Local Potential Approximation, or LPA,
modified minimally to account for the renormalization of the fields~\cite{Tetradis94}, a modification which is usually referred to as LPA'.
  It means that the only momentum dependence of
$\Gamma_k$ is contained in the bare propagator (up to the scaling factor for the field), while the potential part $U_k$ is renormalized:
 \begin{align}
   \Gamma_k = &\int_{\textbf{x},t} \Big\{\Phib \big(\partial_t - \Delta \big) \Phi + \mu \Phib \Delta \Psi\nonumber\\
  & + Z_k \Psib \big(\partial_t - \lambda_k \Delta \big) \Psi + U_k(\Phi+ \Phib,\Psi,\Psib)\Big\}\, . \label{ansatzDPC}
 \end{align}
Furthermore, while some of the results presented below are valid for an arbitrary potential, the integration of the flow and the search 
for a fixed point are performed by Taylor expanding $U_k$ (around a minimum for $\Phi+\Phib$, see appendix \ref{ward-dpc}) to a
 finite order (from third up to sixth order)
 and truncating the corresponding flow equations at the same order. 
  The derivation of the  flow equations is explained in Appendix \ref{nprg}.
The LPA (or LPA') approximations are simple to implement, and straightforward to relate to known perturbative renormalization results.  
 It gives a fairly accurate view of the physics in most problems. However,  it also has limitations, as emphasized in the following.

 For the third order (which corresponds to a renormalization of the coupling constants of the vertices of the DP-C bare action), 
 we find a critical fixed point for any dimensions below $d = 4$ and for any
value of $\mu$. It corresponds to the fixed point precedently described perturbatively in~\cite{Wijland98}.
For $\mu>0$, we also find that a fixed point exists, but the flow cannot reach it since this would require for the vertex
 $\Psi\Psib(\Phi+\Phib)$ to change sign, which
 is prevented in our scheme by the expansion around the minimum. 
 This is manifest in Eq.~(\ref{eqmin}), the corresponding coupling $\tilde{u}_{111}$ cannot vanish.

\begin{figure}
    \centering
    \begin{tikzpicture}
        \node[anchor=south west,inner sep=0] (image) at (0,0) {\includegraphics[width=0.5\textwidth]{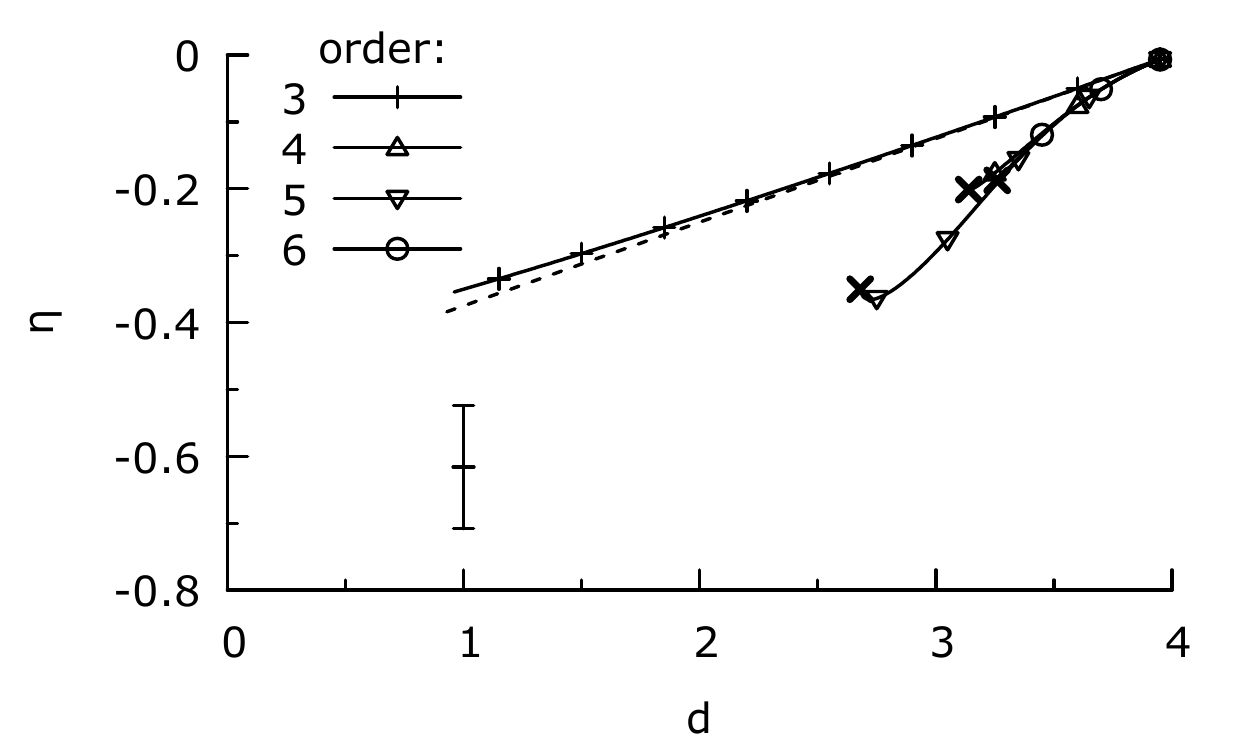}};
    \end{tikzpicture}
    \caption{Values of $\eta$ as a function of the dimension $d$ for $\mu=0$ calculated from the LPA' truncations compared to the one-loop
      result (dashed line)
    and to the simulation~\cite{Dickman07} (isolated point at $d=1$). Thick saltires mark the disappearance of the fixed point. (Colors online)}
    \label{result_mu0}
\end{figure}

From the fourth to sixth order, we find a  fixed point with one unstable direction  for all values of $\mu$, but only for $ d\gtrsim 3$.
 Let us first discuss  the case $\mu=0$.  The values of $\eta$ at the corresponding fixed point are displayed as a function of the dimension  
 in Fig.~\ref{result_mu0}. A striking difference
 is manifest between the lowest and higher orders. For $d=3$, we obtain the value $\eta = -0.3$ at fourth order.
For higher orders, the values of $\eta$ at different dimensions are similar to those of the fourth order, 
but the fixed point solution is lost at a 
dimension slightly higher than $d=3$. If extrapolated to $d=1$, these values seem to be in accordance with the simulations in 
 $d=1$, but the fixed  point solution disappears in dimension $d\simeq 3$.
Unfortunately, it is computationally too costly to implement the seventh and higher orders.
 The way to progress would be to deal with richer ans\"atze,
 for example using the full effective potential without truncation. 
  At this point, as discussed below, it is not clear whether the disappearance of the fixed point is an artifact due
 to the finite-order truncation of the LPA', or if it is related to the intrinsic nature of the DP-C fixed point
  and its relation to DEP.
 
\begin{figure}
    \centering
    \begin{tikzpicture}
        \node[anchor=south west,inner sep=0] (image) at (0,0) {\includegraphics[width=0.5\textwidth]{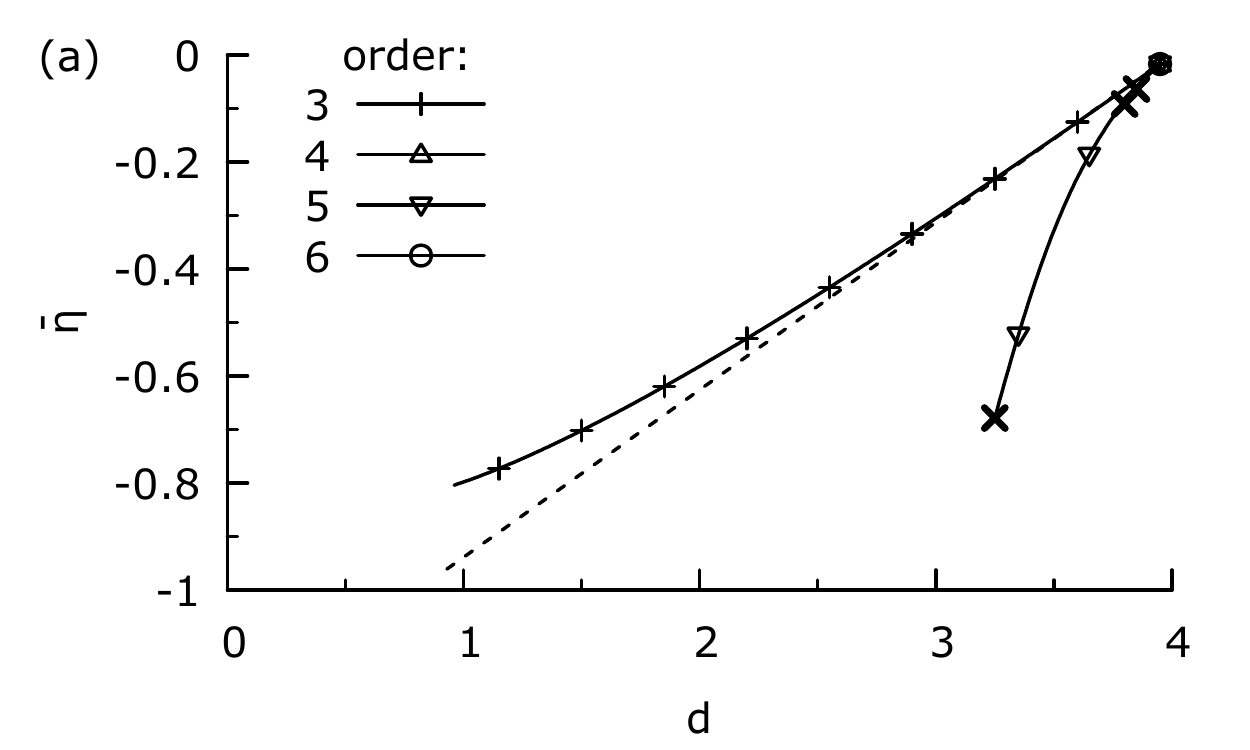}};
        \node[anchor=south west,inner sep=0] (image) at (0,-5) {\includegraphics[width=0.5\textwidth]{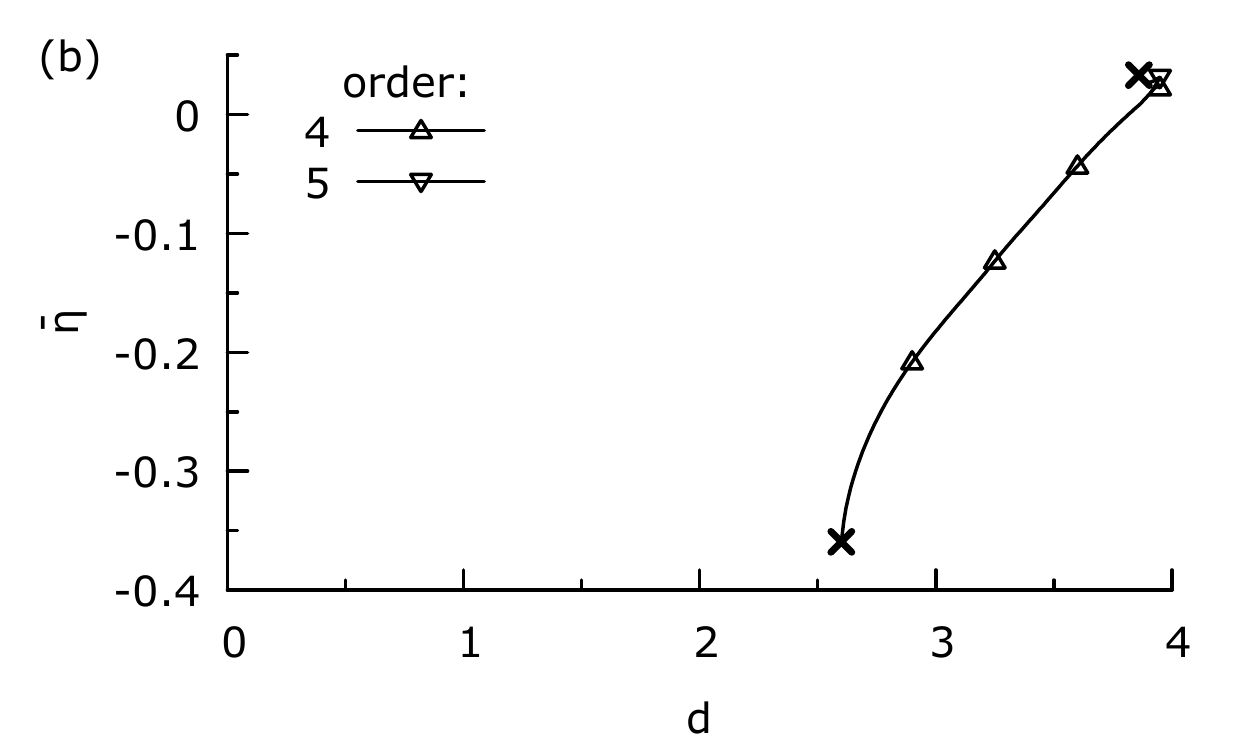}};
    \end{tikzpicture}
    \caption{Values of $\etab$ as a function of the dimension for (a) $\mu=-1$ and (b) $\mu=0.5$ calculated from the LPA' truncations
    compared to the one-loop result (dashed line). Thick saltires mark the disappearance of the fixed point. (Colors online).}
    \label{result_mu}
\end{figure}

In the $\mu<0$ case, this problem seems even sharper (see Fig. \ref{result_mu}), since the domain of existence of the fixed point
 with one unstable direction describing the transition  around $d=4$  is not under control as the truncation order increases. 
 Finally, when $\mu>0$, the same fixed point as for $\mu<0$ is present, but it cannot be reached from the bare action.
 Besides this fixed point,  from the order four in the field expansion, we find a new fixed point in the physical region of parameters.
 In contrast with the  perturbative calculations  but in agreement with  the numerical simulations, 
 this finding supports the existence of a second order phase transition for $\mu>0$.
  However, we are not able to  convincingly assess the convergence of this result with the order of the truncation in the field 
  expansion, since the domain of  existence of this fixed point also   varies significantly
    as the order is increased.
 
Let us note that for $\mu \neq 0$,  as $\eta = 0$, the only non-trivial exponent for the DP-C class is $\bar\eta$
 (see Table \ref{exact}).  This exponent is related to  $\theta'=- \bar \eta / (2 z)$, the critical initial slip exponent \cite{Wijland98}. 
 Hence, it would be interesting to compare the value of $\bar\eta$ obtained for $\mu\neq 0$ in $d=3$
  to simulations in order to test the relevance of the corresponding fixed point. 
 Unfortunately no such determination exists in the literature for $\mu\neq0$.
In both cases, the disappearance of the DP-C fixed point gives rise to the same question as in the $\mu=0$ case.
 Either it is an artifact to be imputed to the finite-order truncation, or it  signals that the transition cannot be 
 described by the DP-C class in low dimensions.

\subsection{RG flow for the DEP action}

Although the LPA' for DP-C shows promising results, the confirmation of the existence of a fixed point with one unstable direction 
driving the absorbing transition in $d=1$ does not seem to be reachable within the kind
of approximation implemented in this paper. Moreover, given that the DP-C fixed point seems to be characterized by a larger symmetry group
   than the DEP one, it is unclear if one should expect the infrared DEP fixed point to be the same as the DP-C one for dimensions
    far from  $d_c$.  On top of this, as already mentioned, some DP-C predictions for critical exponents (such as $\beta=1$ for $\mu \neq 0$)
     seem to be  inconsistent with the results of lattice simulations.
Independently of this last point,
 it is unsatisfactory from a NPRG point of view  not to pinpoint how the DP-C symmetries can emerge along the
flow from a microscopic action lacking these symmetries.

As an attempt to clarify this issue, we studied the flow of a potential $U_k(\Phi,\Phib,\Psi,\Psib)$ symmetric 
under the transformation (\ref{symIII}) related to the DEP action,  Taylor expanded both in $\Phi$ and $\Phib$ around a minimum value.
 We find  that, in dimensions $d\gtrsim3$ where the DP-C fixed point exists,
 the effective action preserves the rescaling symmetry (\ref{symIII}) along the flow, 
 which progressively behaves as the shift symmetry (\ref{symI'}),
  eventually recovered at long distance, for any value of $\mu$, confirming the scenario proposed in Sec. \ref{scenario}.
 However, this scenario only holds for values of $d$ where the DP-C fixed point exists and has the appropriate number of 
 relevant directions in order to control the critical behaviour (which depend on $\mu$ and on the order of the truncation).
 Below this value, the flow leads to another fixed point which is unphysical or diverges.
  This new fixed point is presumably spurious but its existence may be the
 reason why the DP-C fixed point ceases to exist below a certain dimension. We conjecture that if a more elaborate approximation describing 
 a broader theory space than DP-C was used, the same mechanism would occur in small dimensions but the flow would lead to a genuine fixed point,
  which is absent in our truncation.

Let us explain why the LPA'  does not suffice to resolve this issue. 
  First, let us note that the analysis of  the diagrams involved in the perturbative renormalization of the theory shows that the
   conjunction of the $\Psi\Psib$
  term as a multiplicative factor of the potential and the It$\bar{\rm o}$'s prescription~\cite{Canet11} (which forbids tadpoles in the perturbative 
  formalism and also within the LPA' approximation)
 implies that in practice the  $\Phi\Phib$ propagator cannot be renormalized and hence no fixed point other than the DP-C one can be 
 reached starting from the DEP action.
 The same conclusion holds within the LPA' approximation, regardless whether the effective potential is expanded or not.
 In order to study the possible differences between DEP and DP-C, a more elaborate approximation, 
 allowing in particular for a renormalization of the $\Phi\Phib$ propagator  has to be implemented.
In this respect, the Grassberger formulation seems to be a better starting point because, as already mentioned, its variation under 
the time-gauged shift (\ref{symI}) is linear. This may constitute the subject of a future work.

\section{Conclusion}

We analyzed the action of DP-C and DEP, and provided a detailed analysis of their respective symmetries. 
 We then studied the DP-C action in the framework of non-perturbative renormalization,
 using a modified local potential approximation LPA'.  
 We recovered the results of previous field theoretical studies at one-loop in the $\epsilon$-expansion \cite{Wijland98}, 
 that is the existence of a DP-C fixed point for $\mu=0$ and $\mu<0$. 

However, it turns out that this result  cannot be 
 simply extended to $d=1$ and $2$. Indeed, we observed (within the LPA') that this fixed point no longer exists below a certain 
dimension $d\simeq 3$.
  At this stage, two possible explanations can be contemplated. The disappearance of the DP-C fixed point can be an artifact
   of the LPA' approximation implemented in this work, which means that finding the DP-C fixed point is low dimensions requires a 
   higher-order approximation. The alternative 
 interpretation is that the DP-C fixed point ceases to control the absorbing transition in low dimensions, and thus 
 that the transition in the DEP model is driven by another fixed point, which has less symmetries than the DP-C one.
 Investigating this scenario also requires a more elaborate approximation, able to distinguish between the DEP and DP-C renormalizations.
  This interpretation means that the transitions observed in lattice simulations in $d=1$ or 2 do not correspond to 
the DP-C universality class,  but to a genuine DEP class, where the exponent values are not fixed by the exact DP-C identities.
  This scenario could explain the mismatch between the exact predictions coming from the symmetries of the DP-C fixed point,
  and results obtained in lattice simulations. 
 In conclusion, the study of DEP and of its possible coarse-grained counterpart, DP-C, is far from being a settled subject.
We hope to be able  to answer the questions raised in this article in future studies.

\appendix

\section{Ward identities} \label{ward}

Within the NPRG framework, in the presence of the infrared regulator $R_k$, Ward  identities can be derived by considering
 an infinitesimal change of variable in $Z_k$~\cite{Canet15}:
\begin{equation}
 \phi_{\alpha} \to \phi_{\alpha} + \delta_{\epsilon}\phi_{\alpha},\quad \delta_{\epsilon}\phi_{\alpha} = \epsilon(A^{\beta}_{\alpha}\phi_{\beta}+B_{\alpha})\, .
\end{equation}
If the jacobian of the change of variables is one, and using $J_{\alpha} = \delta\Gamma_k/\delta \Phi_{\alpha} + R_k^{\alpha\beta}\Phi_{\beta}$,
one obtains
\begin{equation}
 \delta_{\epsilon}\Gamma_k  = \mean{\delta_{\epsilon}S} +\epsilon R_k^{\alpha\beta}A_{\alpha}^{\gamma}G_{\gamma\beta}\, ,
\end{equation}
where $\delta_{\epsilon}\Gamma_k = (\delta\Gamma_k /\delta \Phi_{\alpha})\delta_{\epsilon}\Phi_{\alpha}$
 and $G_{\alpha\beta}= \delta^2 W_k/( \delta\Phi_{\alpha}\delta\Phi_{\beta})$.
It means that the variation of $\Gamma_k$ is equal to the mean of the variation of $S$ (up to a possible regulator term).

The right-hand side can be made explicit if $\delta_{\epsilon}S$ is linear in the field 
 and the regulator term is invariant. This is a standard result for a global symmetry of the action.
Another  interesting case is when a global shift ($A=0$) symmetry of the action is gauged.
 Under the condition that the momentum dependence of $S$ is only contained in its quadratic part, one can derive a local 
 constraint on $\Gamma_k$. This is the case for the symmetry (\ref{symI'}) of DP-C, detailed below.
 
 \subsection{Ward identities for DP-C and $\nu = 2/d$} \label{ward-dpc}
 The variation of $\Gamma_k$ under the space-time gauged version of (\ref{symI'}) reads
\begin{equation}
 \delta_{\epsilon}\Gamma_k = \delta_{\epsilon}\Big\{\Phib(\partial_t-\Delta)\Phi + \mu\,\Phib\Delta\Psi\Big\}\, .
\end{equation}
This identity means that these two terms are not renormalized, and the remaining part of $\Gamma_k$ must be invariant under 
the transformation (\ref{symI'}). That is, it can only depend on $\Phi+\bar\Phi$, hence the parametrization (\ref{ansatzDPC}).

The exact value for the exponent $\nu$ is then fixed by the duality (\ref{symII}).
Indeed, performing the infinitesimal change of variable $\varphi \to \varphi + \epsilon$ in $Z_k$, one obtains the Ward identity
\begin{equation}
 \int_{{\bf x}, t} \frac{\delta \Gamma_k}{\delta \Phi} = g \frac{\partial  \Gamma_k }{\partial \sigma} \, .
\label{wardnuDPC}
\end{equation}
Evaluating this identity for a uniform and static field configuration then yields
\begin{equation}
 g\,\p_{\sigma} U_k = \p_{\Phi} U_k \, ,\label{wardU}
\end{equation}
where $U_k$ is the effective potential.
Since $U_k$ is analytic, let us consider the expansion of $U_k$ around a minimum $\chi_k$ for $\Phi+\Phib$
\begin{equation}
 U_k(\Phi+\Phib,\Psi,\Psib) = \Psi \Psib \sum_n u_{n,k}(\Psi,\Psib)(\Phi+\Phib-\chi_k)^n\, 
 \label{Ukexp}
\end{equation}
such that $\p_{\Psi\Psib} U_k(\chi_k,0,0) = u_{0,k}(0,0)=0$ (notice that the property $U_k \propto \Psi \Psib$ can be shown to be conserved 
by the flow, which is why the minimum is defined after derivation with respect to $\Psi$ and $\Psib$). 
Let us also note that the identity (\ref{wardU}) is true for any higher order vertex functions evaluated
 in non-zero background field.
Plugging in (\ref{wardU}) the expansion (\ref{Ukexp})  then yields
\begin{equation}
 \forall n\in\mathbb{N},\quad \p_{\sigma}  u_{n,k} = (n+1)u_{n+1,k}\big(\frac{1}{g}+\p_{\sigma}\chi_k\big) \, .
\end{equation}
By evaluating this equation at $\Psi,\Psib=0$, using $u_{0,k}(0,0)=0$ and $u_{1,k}(0,0)\neq0$ as well as the analyticity of the potential,  
one obtains
\begin{align}
 \p_{\sigma}\chi_k = -\frac{1}{g} = cst. \label{ward1}\\
 \forall n\in\mathbb{N},\quad \p_{\sigma}u_{n,k} = 0 \,. \label{ward2}
\end{align}
 (\ref{ward2}) means that $\chi_k$ is the  only relevant  coupling, so that the flow can be
  projected on the critical surface by considering all couplings except $\chi_k$. As usual, the fixed point is
   fully attractive within the critical surface.
Now, if one considers the action detuned from criticality by the infinitesimal quantity $\sigma-\sigma_c$, one has
 $\chi_k \sim \xi^{-d_{\Phi}} \sim (\sigma - \sigma_c)^{\nu\frac{d}{2}}$, where $\xi$ is the correlation length. Combining this consideration with (\ref{ward1}), one gets
\begin{equation}
\nu = \frac{2}{d} \, .
\end{equation}

\subsection{Ward identities for DEP} \label{ward-DEP} 

Let us consider the symmetry (\ref{symIII}). The DEP action (\ref{CSPI}) is invariant uder this transformation, so one is led to the Ward identity
\begin{equation}
\int_{{\bf x}, t} \Big\{ \big(\sqrt{\rho} + \Phi\big)\frac{\delta \Gamma_{\kappa}}{\delta \Phi} - \big(\sqrt{\rho} + \Phib\big)\frac{\delta \Gamma_{\kappa}}{\delta \Phib} + \Psi \frac{\delta \Gamma_{\kappa}}{\delta \Psi} - \Psib \frac{\delta \Gamma_{\kappa}}{\delta \Psib} \Big\}= 0 \, ,
\end{equation}
which translates into the following relation for vertex functions:
\begin{align}
\Big\{\sqrt{\rho}\Big(\frac{\partial}{\partial \phi_0} - \frac{\partial}{\partial \phib_0}\Big) 
+ \phi_0 \frac{\partial}{\partial \phi_0} - \phib_0\frac{\partial}{\partial \phib_0} + \psi_0 \frac{\partial}{\partial \psi_0} - \psib_0 \frac{\partial}{\partial \psib_0} \nonumber\\
+ (a-b+c-d)\Big\}\Gamma_{\kappa}^{(a,b,c,d)}\big(\{\textbf{q}_i,\omega_i\};\phi_0,\phib_0,\psi_0,\psib_0\big)= 0 \, ,
\label{ward-full}
\end{align}
where $a,b,c,d$ are the  respective numbers of functional derivatives with respect to the fields $\Phi, \Phib, \Psi, \Psib$  and the $0$ indice denotes homogeneous background field value.
The counterpart of the duality relation (\ref{symII}) also applies for DEP. By shifting $\varphi$ and $\varphib$ by the same infinitesimal constant in (\ref{CSPI}), one obtains
\begin{equation}
\int_{{\bf x}, t} \Big\{\frac{\delta \Gamma_{\kappa}}{\delta \Phi} + \frac{\delta \Gamma_{\kappa}}{\delta \Phib} \Big\} = \sqrt{\rho} \frac{\partial \Gamma_{\kappa}}{\partial \rho}.\label{wardnuDEP}
\end{equation}
This identity is similar to (\ref{wardnuDPC}), and
 the same derivation  to establish the exact identity for $\nu$  can be used for DEP, leading to
\begin{equation}
 \nu = \frac{1}{d_{\varphi}}\, .
\end{equation}
As emphasized previously, at variance with the DP-C case, the fields $\varphi$ and $\varphib$ 
in the DEP action may be renormalized and acquire an anomalous dimension, such that $\nu$ may differ from $2/d$ 
 at the DEP fixed point.

\section{Implementation of the NPRG for DP-C} \label{nprg}

Within the Derivative Expansion, one focuses on the small momentum and frequency regime of vertex functions. In this regime 
and in the LPA' approximation,  which corresponds to the ansatz (\ref{ansatzDPC}),
the scale-dependence of the effective action is restricted to the potential $U_k$ and to
 the scaling factors $Z_k$ and $\lambda_k$, {\it i.e.} to low order derivatives of propagators at zero momenta and frequencies,
evaluated at a fixed configuration of constant fields. 
In the general case, for two species,   one has to calculate a  $4\times4$
matrix of 2-point correlation functions for $\Psi$, $\Phi$ and their response fields. 
However, for DP-C, the only propagator which can undergo renormalization
in the LPA' approximation is $\Psi\Psib$. As a consequence, only  two scale-dependent numbers 
are needed (on top of potential couplings): $Z_k$, which encode the renormalization of the fields,
  and $\lambda_k$, which gives the scaling between time and space, as encoded in (\ref{ansatzDPC}).
As the $\Phi\Phib$ propagator is not renormalized, $\lambda_k$ does not have a scaling dimension but it can 
still have a sub-dominant dependence on $k$.
 Another approximation is implemented for the potential $U_k$. Instead of treating
  its full field dependence, it is expanded in a polynomial of the fields around the minimum
 $\chi_k$ of $\Phi+\Phib$ defined in Eq. (\ref{Ukexp}), and truncated at a given order.

Let us first define the $\Psi\Psib$ inverse propagator:
\begin{equation}
  \label{gamma2}
  \Gamma_k^{(\Psi\Psib)}({\bf p},\nu) \equiv {\rm FT} \big(\frac{\delta^2 \Gamma_k}{\delta 
  \Psi \delta \Psib}\Big|_{{\rm Min}}\big)\, ,
\end{equation}
where ${\rm Min}$  denotes the static and uniform configuration of the fields  
$\Psi,\Psib \to 0,\, \Phi + \Phib \to \chi_k$ and ``FT''  the
Fourier-transformation, performed after differentiation and evaluation at the minimum configuration.
Furthermore  translational invariance is used so that the left-hand side only depends  on
 one momentum and frequency. The scaling factors  can then be defined as
\begin{align}
  \label{deflambdaZ}
  Z_k &\equiv -i \partial_{\nu} \Gamma_k^{(\Psi\Psib)}({\bf p},\nu)\Big|_{{\bf p},\nu=0}\nonumber\\
  \lambda_k Z_k &\equiv \partial_{p^2} \Gamma_k^{(\Psi\Psib)}({\bf p},\nu) \Big|_{{\bf p},\nu=0}\, ,
\end{align}
and the anomalous dimensions are obtained as
\begin{equation}
  \eta + \etab = - k \partial_k \log Z_k \, .
\end{equation}
 The respective values of each anomalous dimension is provided by the symmetries:
$\eta = 0$ for $\mu \neq 0 $ and $\eta = \etab$ for $\mu = 0$.
 The NPRG flow equation for the inverse propagator reads:

\begin{widetext}
  \begin{equation}
\partial_k \Gamma_k^{(\Psi\Psib)}({\bf p},\nu) =
 \text{Tr}\int_{{\bf q},\omega} \partial_k R_k(q)\cdot G_{k,\epsilon}({\bf q},\omega)\cdot \frac{\partial 
 U_k^{(2)}}{\partial \Psi}\cdot
 G_{k,\epsilon}({\bf p}+{\bf q},\omega+\nu)\cdot \frac{\partial U_k^{(2)}}{\partial \Psib}\cdot G_{k,\epsilon}
 ({\bf q},\omega)\, ,\label{dkgam2}
\end{equation}
\end{widetext}
where $U_k^{(2)}$ is the  4$\times$4 matrix of second derivatives of the potential  with respect to the fields
 $\Phi,\bar\Phi,\Psi$ and $\Psib$
and $G_k$ is the matrix of connected correlation functions, obtained by
inverting the  4$\times$4 matrix $\Gamma_k^{(2)}+R_k$.
The $\epsilon$ subscript account for the It$\bar{\rm o}$'s choice of discetization and indicates that the 
time at which the response field is evaluated is shifted by~$\epsilon$
\begin{equation}
\la \varphib({\bf x}, t) \varphi({\bf x'}, t')\ra_C \to \la \varphib({\bf x}, t + \epsilon) \varphi({\bf x'}, t')\ra_C\, .
\end{equation}
In Fourier space, this leads for example to
\begin{equation}
  G^{\varphi\varphib}_{k,\epsilon}(q,\omega) = e^{i \epsilon \omega} G^{\varphi\varphib}_k(q,\omega)\, ,
  \label{ito}
\end{equation}
see~\cite{Canet11} for the justification of this procedure. 
 From Eq.~(\ref{dkgam2}), one then deduces the flow equations for $Z_k$ and $\lambda_k$ 
 following the definitions (\ref{deflambdaZ}).
 Note that additional tadpole diagrams (proportional to fourth derivatives of the potential), which vanish within the LPA'  
 in It$\bar{\rm o}$'s convention,  are not included in  Eq.~(\ref{dkgam2}).
 
The flow equation for the potential reads
\begin{align}
  \partial_k U_k
    =& \int_{{\bf q}, \omega}
  \Big\{
    \partial_k  q^2 r(\frac{q^2}{k^2}) G^{\varphi\varphib}_{k,\epsilon}(q,\omega)\nonumber\\
   &+ \partial_k ( \lambda_k Z_k  q^2 r(\frac{q^2}{k^2})) G^{\psi\psib}_{k,\epsilon}(q,\omega)
    \Big\}\, .
\end{align}
With  It$\bar{\rm o}$'s shift convention  (\ref{ito}), the integration on $\omega$ amounts to a 
sum on the residues of $G$ in the upper-half plane, operation which is denoted by ${\rm Res}_+$ in the following.
Using the ansatz (\ref{ansatzDPC}) for DP-C one obtains
\begin{widetext}
\begin{align}
 \partial_k U_k = \int_{{\bf q}}
 i {\rm Res}_+ \Big\{
  q^2 \partial_k r \frac{ \big( i \omega + q^2 (1+r) + U_k^{(2,0,0)} \big) P_{\Psi }+ \mu q^2 \Big(U_k^{(1,0,1)} \big( - Z_k i \omega + h  \big) - U_k^{(0,0,2)} U_k^{(1,1,0)} \Big) + Q}
          {\Delta}\nonumber\\
+
  q^2 \partial_k ( \lambda_k Z_k r) \frac{  \big(i \omega  Z_k + h\big)P_{\Phi } + \mu  q^2 U_k^{(1,0,1)} \big(-i \omega + q^2 (1+r)\big)-2 q^2 (1+r) U_k^{(1,0,1)} U_k^{(1,1,0)}}{\Delta}
\Big\}\, ,
\end{align}
\end{widetext}
where $r = r(q^2/k^2)$ is the adimensionned regulator defined in (\ref{Theta}),
\begin{equation}
U_k^{(i,j,l)} =  \frac{\partial^{i+j+l} U_k }{\partial(\Phi+\Phib)^i \partial\Psi^j \partial\Psib^l}\, ,
\end{equation}
 and with the following shorthand notations:
\begin{align} 
  h = &\lambda_k  Z_k q^2 (1+r) + U_k^{(1,1,0)}\nonumber\\
  Q = &U_k^{(0,2,0)} \big(U_k^{(1,0,1)}\big)^2+U_k^{(0,0,2)} \big(U_k^{(1,1,0)}\big)^2 \nonumber\\
      &- 2 h U_k^{(1,1,0)} U_k^{(1,0,1)}\nonumber\\
  P_{\Phi} = & \omega ^2 + q^2 (1+r) \big(q^2 (1+r) + 2 U_k^{(2,0,0)}\big)\nonumber\\
  P_{\Psi} = &\big( Z_k\omega \big)^2 + h^2-U_k^{(0,0,2)} U_k^{(0,2,0)}\nonumber\\
\Delta = & P_{\Psi } P_{\Phi } - 2 Z_k U_k^{(1,0,1)} \mu  q^2 \omega ^2 + 2 q^2 Q (1+r) \nonumber\\
&+ 2 \mu  q^4 (1+r) \big(h U_k^{(1,0,1)}-U_k^{(0,0,2)} U_k^{(1,1,0)}\big)
\nonumber\\
&+\mu ^2 q^4 \big((U_k^{(1,0,1)})^2-U_k^{(0,0,2)} U_k^{(2,0,0)}\big)\, .
\end{align}

By differentiating $U_k$ and evaluating the result at the minimum configuration defined in
(\ref{gamma2}), one is left with only two possible poles: $i q^2 (1 + r)$ and $i \lambda_k q^2 (1 + r)$. More precisely, the integrand is now a rational function of 
\begin{align}
\gamma &= i \omega + q^2(1+r)\nonumber\\
\gamma_{\lambda} &= i \omega + \lambda_k q^2(1+r)\, ,
\end{align}
and their respective complex conjugate. 
The non-zero contributions to the integral arise from one or more $\gamma_{(\lambda)}$ factors
 in the denominator. With the notation
\begin{equation}
u_{ijl} \equiv \frac{1}{i!j!l!}\frac{\partial^{i+j+l} U_k }{\partial(\Phi+\Phib)^i \partial\Psi^j 
\partial\Psib^l}\Big|_{\Psi,\Psib \to 0, \Phi + \Phib \to \chi_k}\, ,
\end{equation}
one obtains for example for $u_{011}$
\begin{widetext}
\begin{align}
 \partial_k u_{011} =& \int_{{\bf q}}
 i {\rm Res}_+ \Bigg\{
  q^2 \partial_k r \Big[\frac{q^2 u_{111}^2 \big(\mu -2 \lambda  (r+1) Z\big)-2 Z \gamma _{\lambda }^* \big(u_{211} Z \gamma _{\lambda }^*+\mu  q^2 u_{112}\big)}{Z^2 \big(\gamma _{\lambda }^*\big){}^2 \big(\gamma
   ^*\big)^2}\nonumber\\
&+\frac{2 u_{111} \big(\mu  q^2 u_{012} \gamma _{\lambda }^*+2 \lambda ^2 q^4 (r+1)^2 u_{111} Z\big)}{\gamma _{\lambda } Z^2 \big(\gamma _{\lambda }^*\big)^2 \big(\gamma ^*\big)^2}\Big]
+ q^2 \partial_k ( \lambda_k Z r) \Big[
    \frac{2 q^2 u_{111} \big(2 \mu  u_{012} \gamma_{\lambda }^*+\big(\lambda^2-1\big) q^2 (r+1)^2 u_{111} Z\big)}{\gamma  (\lambda -1) Z^3 \big(\gamma_{\lambda}^*\big)^3 \gamma^*}\nonumber\\
&+\frac{4 u_{012} \big((\lambda -1)
   u_{021} \gamma ^*- \mu  q^2 u_{111}\big)}{(\lambda -1) \gamma_{\lambda } Z^3 \big(\gamma_{\lambda }^*\big)^2 \gamma^*} -
2\frac{Z\gamma_{\lambda }^* \big(u_{121} \gamma ^*+\mu  q^2 u_{112}\big)+ q^2 u_{111}^2 \big((r+1) Z-\mu \big)}{Z^3 \big(\gamma _{\lambda }^*\big)^3 \gamma ^*}
         \Big]\Bigg\}\nonumber\\
=\int_{{\bf q}}\Bigg\{&q^2 \partial_k ( \lambda Z r) \frac{ \lambda ^2 (r+1) u_{111}^2 Z+u_{012} \big((\lambda +1)^2 (r+1) u_{021}+ (2 \lambda +1) \mu  u_{111}\big)}{ \lambda ^2 (\lambda +1)^2 q^4 (r+1)^3 Z^3}
+q^2 \partial_k r \frac{u_{111} \big( \lambda  (r+1) u_{111} Z+\mu  u_{012}\big)}{\lambda  (\lambda +1)^2 q^4 (r+1)^3 Z^2}
 \Bigg\}\, ,
\end{align}
\end{widetext}
where the dependence in $k$ has been omitted for notational simplicity.
At this point, the procedure is the same as for  NPRG  applied to equilibrium systems.
 With our choice of regulator, the momentum integral can be performed analytically and, following the previous example,
one is left with
\begin{widetext}
\begin{align}
 k \partial_k u_{011} = 8 k^{d-2} v_d \Big\{ \frac{ \mu  u_{012} u_{111} \big[\lambda  \big(8 + 2 d -\etab +2 k \partial_k\lambda-\eta\big)+2 \lambda ^2 \big(4 + d -\etab-\eta\big)+k \partial_k\lambda\big]}{(d+2) (d+4)\lambda ^2 (\lambda +1)^2 Z^2}+\nonumber\\
\frac{ u_{012} u_{021} \big[\lambda  \big(2 + d -\etab-\eta\big)+k \partial_k\lambda\big]}{d (d+2) \lambda ^2 Z^2}+
\frac{ u_{111}^2 \big[\lambda  \big(2 + d -\etab-\eta \big)+d+k \partial_k\lambda+2\big]}{d (d+2) (\lambda +1)^2 Z}\Big\}\, ,
\end{align}
\end{widetext}
where $d$ is the spatial dimension and $v_d = 1/(2^{d+1}\pi^{\frac{d}{2}}\Gamma(\frac{d}{2}))$.
The last step is to introduce rescaled variables
\begin{align}
  u_{ijl} &=
  \begin{cases}
    Z^{\frac{j+l}{2}} k^{2+d-(i+j+l)\frac{d}{2}} \tilde u_{ijl}, & \text{if } \mu = 0 \\  
    Z^l k^{2+d-(i+j+l)\frac{d}{2}} \tilde u_{ijl}, & \text{otherwise}
  \end{cases}\nonumber\\
\chi &= k^{\frac{d}{2}}\tilde \chi\nonumber\\
s &= \ln{\frac{\Lambda}{k}}\, .
\end{align}

 The flow equation of the rescaled minimum is deduced from the implicit equation
 $\p_{\psi\bar \psi} U_k\big|_{\rm Min} =0$ in terms of the rescaled variables.
 For $\mu=0$, one obtains
\begin{align}
  \partial_s \tilde \chi =& \frac{d}{2} \tilde \chi + 8 v_d \Big\{\frac{\tilde{u}_{111} (2 + d + \lambda  (2 + d-2 \eta)- \partial_s \lambda)}{d (d+2) (\lambda +1)^2}\nonumber\\
  &-\frac{\tilde{u}_{021}^2 (\lambda  (2 + d-2 \eta)-\partial_s \lambda)}{d (d+2) \lambda ^2
   \tilde{u}_{111}}\Big\}\, .
   \label{eqmin}
\end{align}
In the case $\mu\neq 0$, one obtains
\begin{align}
  \partial_s \tilde \chi& = \frac{d}{2} \tilde \chi + 8 v_d\Big\{\frac{\tilde{u}_{111} \big(2 + d + \lambda  \big(2 + d - \etab\big)-\partial_s \lambda\big)}{d (d+2) (\lambda +1)^2}
  \nonumber\\
 &+ \frac{\mu  \tilde{u}_{012} \big(2 (d+4) \lambda  (\lambda +1)-(2 \lambda + 1)(\partial_s \lambda + \lambda \etab)\big)}{(d+2) (d+4) \lambda ^2 (\lambda
   +1)^2}\nonumber\\
   &+\frac{\tilde{u}_{021} \tilde{u}_{012} \big(\lambda  \big(2 + d-\etab\big)-\partial_s \lambda\big)}{d (d+2) 
   \lambda ^2 \tilde{u}_{111}}\Big\}\, .
\end{align}

To get the full set of coupled ordinary differential equation for $\tilde \chi$, $\lambda$ and 
the coefficients $\tilde u_{ijl}$ up to the truncation order, the above procedure is 
 systematically implemented
 using Mathematica. We verified that $\tilde \chi$ decouples from the other couplings, which implies that
 it is an unstable direction of the flow with eigenvalue $d/2$ and that, once its flow is excluded, the flow of
  the other couplings is fully attractive (if a critical fixed point exists).
   In consequence, the integration of the flow can be simply done numerically, without the need to 
   perform any fine-tuning.

\end{document}